\documentclass[twocolumn,showpacs,prl,amsmath,amssymb]{revtex4}
\usepackage{graphicx}
\usepackage{dcolumn}
\usepackage{bm}

\usepackage{color}

\begin{document}

\title{Splitting of ISGMR strength in the light-mass nucleus $^{24}$Mg due to ground-state deformation}

\author{Y. K. Gupta}
\thanks{Permanent address: Bhabha Atomic Research Centre, Mumbai 400085, India} 
\affiliation{Physics Department, University of Notre Dame, Notre Dame, IN 46556, USA}

\author{U. Garg}, 
\affiliation{Physics Department, University of Notre Dame, Notre Dame, IN 46556, USA}
\author{J. T. Matta, D. Patel, T. Peach}
\affiliation{Physics Department, University of Notre Dame, Notre Dame, IN 46556, USA}

\author{J. Hoffman}
\thanks{Present address: Volcano Corporation, San Diego, CA 92130, USA}
\affiliation{Physics Department, University of Notre Dame, Notre Dame, IN 46556, USA}

\author{K. Yoshida}
\affiliation{Graduate School of Science and Technology, Niigata University, Niigata 950-2181, Japan}
\affiliation{Center for Computational Sciences, University of Tsukuba, Tsukuba 305-8577, Japan}

\author{M. Itoh} 
\thanks{Present address: Cyclotron and Radioisotope Center, Tohoku University, Sendai 980-8578, Japan}
\author{M. Fujiwara}
\author{K. Hara}
\author{H. Hashimoto}
\author{K. Nakanishi}
\author{M. Yosoi}
\affiliation{Research Center for Nuclear Physics (RCNP), Osaka University, Osaka 567-0047, Japan}
\author{H. Sakaguchi, S. Terashima, S. Kishi, T. Murakami}
\author{M. Uchida}
\thanks{Present address: Department of Physics, Tokyo Institute of Technology, Tokyo 152-8850, Japan}
\author{Y. Yasuda}
\affiliation{Department of Physics, Kyoto University, Kyoto 606-8502, Japan}

\author{H. Akimune}
\affiliation{Department of Physics, Konan University, Hyogo 658-8501, Japan}

\author{T. Kawabata}
\thanks{Present address: Department of Physics, Kyoto University, Kyoto 606-8502, Japan}
\affiliation{Center for Nuclear Study, University of Tokyo, Wako, Saitama 351-0198, Japan}

\author{M. N. Harakeh}
\affiliation{KVI-CART, University of Groningen, 9747 AA Groningen, The Netherlands}

\date{\today}

\begin{abstract}
The isoscalar giant monopole resonance (ISGMR) strength distribution in $^{24}$Mg  has been determined  from background-free  inelastic scattering of 386-MeV $\alpha$ particles at extreme forward angles, including 0$^{\circ}$. 
The ISGMR strength distribution has been observed for the first time to have a two-peak structure in a light-mass nucleus. This splitting of ISGMR strength is explained well by microscopic theory in terms of the prolate deformation of the ground state of $^{24}$Mg.
\end{abstract}


\maketitle

The isoscalar giant monopole  resonance (ISGMR) has been investigated in a wide range of atomic nuclei from $^{12}$C to $^{208}$Pb 
\cite{Itoh_32S,Lui_PRC_1981,BJohn2003, Li_PRL2007, Li_2010, Darshna2014, Lui_16O_2001} and has been shown to be an effective way to obtain an experimental value for the nuclear incompressibility \cite{Harakeh_book, colo_garg_sagawa}.  However, identification of the full $E0$ energy-weighted sum rule (EWSR) in lighter nuclei ($A<60$) has not been possible due to fragmentation of the strength, the nearly complete overlap of the ISGMR with the isoscalar giant quadrupole resonance (ISGQR) and other multipoles, uncertainties in the extraction of the strength distributions, and the difficulty in distinguishing the multipole strength from other direct processes (quasi-free knock-out process, for example). The fragmented  ISGMR strength in lighter nuclei further renders it nearly impossible to identify effects such as the theoretically predicted splitting of the ISGMR due to ground-state deformation. While the splitting of the isovector giant dipole resonance (IVGDR) due to deformation has been documented in a number of nuclei \cite{Harakeh_book}, a similar effect on the ISGMR strength has been reported so far only in the deformed Sm nuclei\cite{garg_prl_1980,tamu1,YB_154Sm_2004,Itoh_prc2003} and in the fission decay of $^{238}$U \cite{kvi_prl}; this ``ISGMR splitting'' is understood  in terms of the mixing of the ISGMR with the $K^{\pi}$=0$^{+}$ component of the ISGQR \cite{garg_prl_1980}.  

Recent microscopic calculations \cite{Yoshida_2010_24Mg,Peru2008} in the deformed Hartree-Fock-Bogoliubov (HFB) approach and the quasiparticle random-phase approximation (QRPA)
with a Skyrme and Gogny energy-density functional have shown that the ISGMR strength distribution exhibits a two-peak structure due to deformation even in light-mass nuclei. In particular, the calculations indicate that the prolate-deformed ground state of $^{24}$Mg leads to a two-peak ISGMR strength structure because of the aforementioned mixing of the ISGMR with the $K^{\pi}$=0$^{+}$ component of the ISGQR. Experimentally, the isoscalar giant resonance strength in $^{24}$Mg has been investigated using inelastic scattering of $\alpha$ particles  and $^{6}$Li  \cite{YB_24Mg_1999, YB_24Mg_6Li, YB_24Mg_2009, Kawabata_24Mg_2013}. These measurements have shown excitation of similar $E0$ EWSR strengths. However, none has revealed any discernible ``ISGMR splitting'' due to deformation effects.

In this Letter, we report the first experimental evidence of a two-peak structure in the ISGMR strength distribution in the light-mass nucleus $^{24}$Mg, as obtained from ``background-free'' inelastic $\alpha$-scattering spectra obtained at extreme forward angles, including 0$^{\circ}$. 
The experimentally observed peak positions and giant resonance strengths are in good agreement, within the experimental uncertainties, with the calculated ISGMR strength distributions based on a  prolate ground-state deformation for $^{24}$Mg.

Inelastic scattering of 386-MeV $\alpha $ particles
was measured at the ring cyclotron facility
of the Research Center for Nuclear Physics (RCNP), Osaka
University. A self-supporting  foil (0.7 mg/cm$^{2}$) of enriched ($>99\%$) $^{24}$Mg  was  employed as the target. 
Inelastically scattered $\alpha$ particles were momentum analyzed with the high-resolution magnetic spectrometer ``Grand
Raiden" \cite{Fujiwara_GR}, and the  horizontal and vertical  positions of
the $\alpha$ particles were measured using a focal-plane detector
system composed of two position-sensitive multiwire drift
chambers (MWDCs) and two scintillators \cite{Itoh_prc2003}. The MWDCs
and scintillators enabled us to achieve particle identification and
to reconstruct the trajectories of the scattered particles.  The
vertical-position spectrum obtained in the double-focusing
mode of the spectrometer was exploited to eliminate the
instrumental background \cite{Uchida_PLB2003,Itoh_prc2003}.

\begin{figure}
\centering\includegraphics [trim= 0.11mm 0.5mm 0.1mm 0.1mm,
angle=360, clip, height=0.18\textheight]{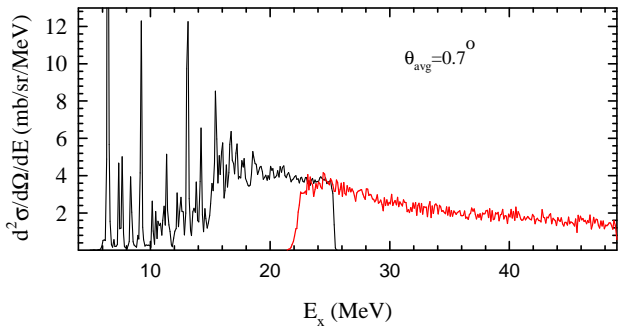}
\caption{(color online) Excitation-energy spectra for $^{24}$Mg($\alpha, \alpha'$) at an
averaged spectrometer angle, $\theta_\mathrm{avg}$ = 0.7$^{\circ}$. The black and 
grey (red)  lines show the energy spectra obtained from the low-$E_{x}$ and high-$E_{x}$
measurements, respectively.}
\label{Low_High}
\end{figure}

Data for elastic scattering and inelastic scattering to the low-lying states were taken in the angular range of 3.5$^{\circ }$ to 26.5$^{\circ }$.
Giant resonance measurements  were performed  at very forward central angles of the spectrometer (from 0$^{\circ }$ to 10.4$^{\circ }$). 
Using the ray-tracing technique, the angular width  of 1.8$^{\circ }$ for each central angle was divided into five equal regions. Measurements were made for two magnetic-field settings of  Grand Raiden,  resulting in spectra covering excitation energies from about  4 to 27 MeV and from 24 to 50 MeV--the low-$E_{x}$ and the high-$E_{x}$ spectra, respectively. Data were also taken with a $^{12}$C target at each setting of central angle and magnetic field  of the spectrometer, providing a precise energy calibration. The high-$E_{x}$  spectrum connects smoothly with the low-$E_{x}$ one, as demonstrated in Fig. \ref{Low_High} for an averaged spectrometer angle, $\theta_\mathrm{avg}$=0.7$^{\circ }$.

\begin{figure}[h!]
\centering\includegraphics [trim= 0.11mm 0.5mm 0.1mm 0.1mm,
angle=360, clip, height=0.3\textheight]{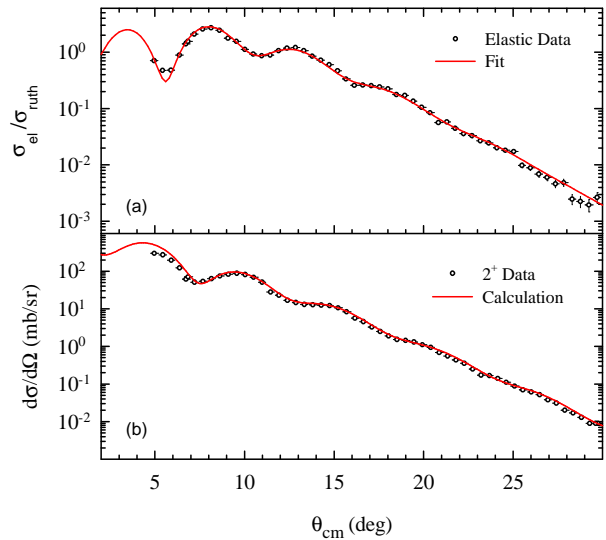}
\caption{(color online) (a) Angular distribution of the ratio of the differential
cross sections for elastic scattering to Rutherford scattering of
386-MeV $\alpha$ particles off $^{24}$Mg. The solid (red) line is the optical-model fit to the data. (b) Angular distribution of differential
cross sections for the 1.368-MeV 2$^{+}$ state. The solid (red) line
shows the result of the DWBA calculation (see text).}
\label{Elastic_2p}
\end{figure}

Elastic scattering angular distributions were used to extract the optical-model parameters (OMPs) for the ``hybrid" potential proposed by Satchler and Khoa \cite{Satchler_Khoa1997}. In this procedure, the real part of the optical potential is generated by single-folding with a density-dependent Gaussian $\alpha$-nucleon interaction \cite{Li_2010}.  The computer codes SDOLFIN and DOLFIN \cite{DOLFIN}  are used to calculate the shape of the real part of the potential and the form factors, respectively. Radial moments for $^{24}$Mg are obtained by numerical integration of
the Fermi mass distribution assuming $c$=3.0453 fm and $a$=0.523 fm \cite{Fricke1995}. A Woods-Saxon form was used for the imaginary part of the optical potential. The imaginary potential parameters ($W$, $R_{I}$, and $a_{I}$), together
with the depth of the real part ($V$) are obtained  by fitting
the elastic-scattering cross sections using the computer code PTOLEMY \cite{ptolemy1,ptolemy2}.  The best fit to the elastic cross-section data (normalized to the Rutherford cross section) obtained from a $\chi^{2}$ minimization is presented in Fig. \ref{Elastic_2p}(a). The OMPs thus determined are given in Table \ref{tab:table1}. 

\begin{table}
\caption{\label{tab:table1} Optical-model parameters obtained by fitting elastic scattering
data. Also listed is the $B(E2)$ value for the 1.368-MeV 2$^{+}$ state from Ref. \cite{BE2_24Mg}.}
\begin{ruledtabular}
\begin{tabular}{cccccc}
  V     & W        & $R_{I}$      &$a_{I}$ &$R_{C}$    &$B(E2)$ \\
   (MeV)&   (MeV)  &    fm       & fm       &fm          & e$^{2}$b$^{2}$ \\
\hline \\
33.1     &  36.1   &  3.87      & 0.778     & 3.04   &0.0432\\\

\end{tabular}
\end{ruledtabular}
\end{table}

Using the known $B(E2)$ value from the literature (also listed in Table \ref{tab:table1}) and the OMPs thus obtained, the angular distribution for the 1.368 MeV 2$^{+}$ state was calculated in the distorted-wave Born Approximation (DWBA) framework. An excellent agreement
between the calculated and experimental angular distributions  for the 2$^{+}$ state, as shown in Fig. \ref{Elastic_2p}(b), establishes the appropriateness of the OMPs. 

\begin{figure}[h!]
\centering\includegraphics [trim= 0.11mm 0.5mm 0.1mm 0.1mm,
angle=360, clip, height=0.45\textheight]{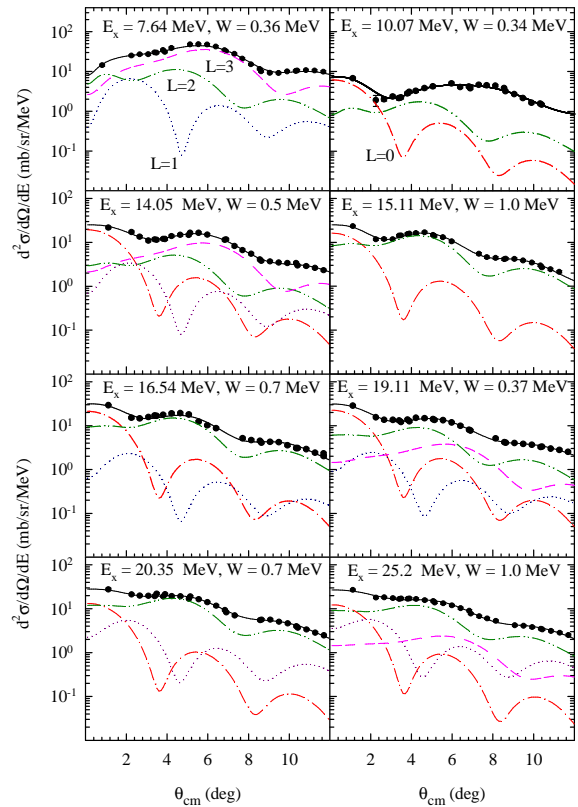}
\caption{(color online) Representative angular distributions of inelastic $\alpha$
scattering from $^{24}$Mg. The solid line (black) through the data shows the sum of various multipole
components obtained from MDA. The dash-dotted (red), dotted (blue), dash-dot-dotted (green), and dashed (pink) curves
show contributions from $L$ = 0, 1, 2, and 3, respectively, with the transferred angular momentum $L$ specified
along with the curves. The mean $E_{x}$ value, as well as the bin width $W$, are also provided for each case.}
\label{MDA}
\end{figure}

The inelastic-scattering cross sections were divided into energy bins of different sizes. For the $E_{x}$ region from 4 to 20 MeV, the size of the bin was chosen to accommodate the discrete peaks. Because the discrete structure of the strength distribution diminishes for $E_{x}>$21 MeV (see Fig. \ref{Low_High}), the bin size in this energy domain was chosen to be 1 MeV to reduce statistical fluctuations. 
The laboratory angular distribution for each excitation-energy bin was converted to the center-of-mass frame using the standard Jacobian and relativistic kinematics. Representative angular distributions are shown in Fig. \ref{MDA}. 

The experimental angular distributions thus obtained consist of contributions from various multipoles. A multipole-decomposition analysis (MDA) was carried out to disentangle these different contributions \cite{Li_2010}. In this process, the experimental double-differential cross sections are
expressed as linear combinations of calculated DWBA double-differential cross sections for different
multipoles as follows:
\begin{equation}
\frac{d^{2}\sigma ^{\mathrm{exp}} (\theta_{\mathrm{c.m.}}, E_{x})}{d\Omega dE}= \sum\limits_{L=0}^6 a_{L}(E_{x})\frac{d^{2}\sigma_{L}^{\mathrm{DWBA}}(\theta_{\mathrm{c.m.}}, E_{x}) }{d\Omega dE} 
\end{equation}
where $a_{L}(E_{x})$ is EWSR fraction for the $L^{th}$ component. $\frac{d^{2}\sigma_{L}^{\mathrm{DWBA}} }{d\Omega dE} (\theta_{\mathrm{c.m.}}, E_{x})$ is the calculated DWBA cross section corresponding to 100\% EWSR for the L$^{th}$ multipole. We used transition densities and sum rules for various multipolarities as described in Refs. \cite{Harakeh_book,Satchler1987,Harakeh1981}. 
The $a_{L}(E_{x})$ are determined from the $\chi^{2}$ minimization technique, with the uncertainties estimated by changing the magnitude of one component, $a_{L}(E_{x})$, until refitting by varying the other components resulted in an increase in $\chi^{2}$ by 1 \cite{YB_24Mg_1999, YB_24Mg_2009, Itoh_prc2003}. Fits from the MDA, including contributions from $L\leq$3 modes, are shown in Fig. \ref{MDA}.

\begin{table}[h!]
\begin{ruledtabular}
\caption{\label{tab:table2} $B(EL)$ values (or the percentage of the total EWSR, where appropriate) 
obtained for some low-lying states in $^{24}$Mg.}
\begin{tabular}{lcll}

  E$_{x}$     & $J^{\pi}$          &$B(EL)$ (This work)                &$B(EL)$   (Ref. \cite{YB_24Mg_1999})  \\
   (MeV)       &                         & (e$^{2}$b$^{L}$)                    & (e$^{2}$b$^{L}$)\footnotemark[1]    \\
\hline
4.17  &2$^{+}$  & 3.36(9)$\times$10$^{-3}$                            & [3.48$\times$10$^{-3}$]\footnotemark[2]    \\
6.04  &4$^{+}$  & 2.92(6)$\times$10$^{-4}$                            & 2.3$^{+0.7}_{-1.0}\times$10$^{-4}$                 \\
6.41  &0$^{+}$  & 4.85(12)\%                                                  &5.6(1.0)\%    \\
7.33  &2$^{+}$  & 1.45$^{+0.01}_{-0.06}\times$10$^{-3}$      &1.5$^{+0.8}_{-0.2}\times$10$^{-3}$            \\
7.64\footnotemark[3] &1$^{-}$  & 3.1(6)\%                                                      &3(1)\%     \\
        &3$^{-}$  & 1.06(3)$\times$10$^{-3}$                        & 1.1$^{+0.2}_{-0.3}\times$10$^{-3}$       \\
8.41  &3$^{-}$   & 1.46$^{+0.06}_{-0.15}\times$10$^{-3}$      &  2.5(4)$\times$10$^{-3}$                \\
 \end{tabular}
 \end{ruledtabular}
 \footnotetext[1]{Ref. \cite{YB_24Mg_1999} lists the units as e$^{L}$fm$^{2L}$ for the numbers given here; we believe it was a typographical error.}
 \footnotetext[2]{Value determined from deformation parameters obtained from Ref. \cite{KVI_1981_BE2} by using implicit folding procedure
 \cite{Harakeh_book}.}
 \footnotetext[3]{Ref.  \cite{YB_24Mg_1999} reports two peaks at 7.555 MeV (1$^{-}$) and 7.616 MeV (3$^{-}$), respectively. They are not resolved in this work.}
\end{table}

In Table \ref{tab:table2}, $B(EL)$ values (or the \%EWSR) obtained in the present measurement for several discrete states are compared with those reported in the literature. Generally close agreement with the previous results establishes the reliability of the MDA procedure.

\begin{figure}[t!]
\centering\includegraphics [trim= 0.11mm 0.5mm 0.1mm 0.1mm,
angle=360, clip, height=0.25\textheight]{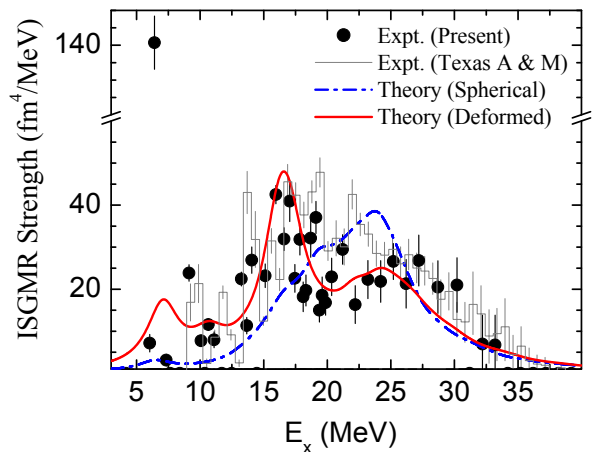}
\caption{(color online) ISGMR strength distribution in $^{24}$Mg extracted in the present work (solid circles). The dash-dotted (blue)  and solid (red) lines  show microscopic calculations for spherical and prolate ground-state deformation, respectively. Also, shown is the ISGMR strength distribution from the Texas A \& M work (grey histograms) \cite{dhy_priv}.}
\label{ISGMR}
\end{figure}

Experimentally determined ISGMR strength distribution, the focus of this paper, is presented in Fig. \ref{ISGMR}. The extracted strength distributions for other multipoles will be presented and discussed in detail in a forthcoming paper \cite{guptanext}.

The ISGMR distribution consists of a two-peak structure: a narrow peak at $E_x\sim$16 MeV, and a broad peak at $E_x\sim$24 MeV. A total of 57$\pm$7\% $E0$ EWSR is exhausted over the excitation-energy region 6--35 MeV.
The two-peak structure is 
very similar to the ISGMR distribution observed in $^{154}$Sm \cite{tamu1,YB_154Sm_2004,Itoh_prc2003}, thus being strongly indicative of resulting from the deformation of the ground state. A comparison of the experimental and theoretical strength distributions further establishes that this structure corresponds to that of a deformed nucleus.
Incidentally, the  extracted ISGMR strength distribution is generally similar to that reported by the Texas A \& M group \cite{YB_24Mg_6Li, YB_24Mg_2009, dhy_priv}, except that the two-peak structure was not as directly discernible there as it is in the present work.

The theoretical strength distribution was obtained 
as a self-consistent solution of the deformed HFB and QRPA equations employing the Skyrme SkM* functional \cite{Bartel1982}.
Details of the calculation scheme can be found in Refs. \cite{Yoshida2008,Yoshida2013}. 
In the present calculations, the smearing width of 3 MeV was introduced to take into account the spreading effects. The SkM* functional gives an intrinsic quadrupole moment $Q_0 =$ 54.0 $e$fm$^2$, which is consistent 
with the value 65.9 $e$fm$^2$ estimated from the measured $B(E2)$ of the first $2^{+}$ state listed in Table~\ref{tab:table1} assuming the nucleus as a rigid rotor. 
In the energy region of 6 to 35 MeV, the obtained IS monopole strength exhausts $83\%$ of EWSR.  
Thus, the theoretical strengths have been scaled down by a factor 0.57/0.83 = 0.69
in Fig.~\ref{ISGMR} for comparison with the experimental data. 
This apparent mismatch between theoretical and experimental strengths is not particularly worrisome considering that the experimental strengths can have up to $\sim$20\% uncertainty resulting from the choice of the OMPs used, and from the DWBA calculations, as has been noted in previous work as well \cite{Harakeh_book,Li_2010}.
In addition to the strength obtained for the prolate-deformed ground state, 
 the strength distribution obtained for a spherical configuration is also shown in Fig.~\ref{ISGMR}.
The peak around 16 MeV appears only when the ground state is deformed.
Emergence of this lower peak in the calculations is due to coupling to the $K^{\pi}=0^+$ component of the ISGQR. This direct comparison of the experimental data with the expected strength distributions for the spherical and deformed ground state for $^{24}$Mg is critical in establishing the observed ISGMR strength as corresponding to that from the deformed ground state. We also note that the QRPA approach employing the Gogny D1S effective force \cite{Peru2008} also leads to a two-peak structure of the ISGMR strength (with peaks around 18 MeV and 25 MeV) due to the ground-state deformation; however, no results for the spherical case were provided in that work. 

In summary, we have measured the isoscalar monopole strength distribution in the light nucleus $^{24}$Mg via small-angle inelastic scattering of $\alpha$ particles. Instead of the generally expected broad fragmentation of the ISGMR strength, we observe a two-peak structure which results from the deformation of the ground state. The observed strength distribution is in good agreement with microscopic calculations for a prolate-deformed ground state for $^{24}$Mg and is in contrast with that expected if a spherical ground state is assumed for this nucleus. This is the first time that the splitting of the ISGMR has been observed in a very light nucleus.

The authors acknowledge the efforts of the staff of the RCNP Ring Cyclotron Facility in providing a high-quality, halo-free $\alpha$-particle beam required for the challenging measurements reported in this paper. Discussions with Prof. G. Col\`{o} on the theoretical aspects of this work are also gratefully acknowledged. This work has been supported in part by the U.S. National Science Foundation (Grant Numbers INT-9910015, PHY04-57120, and PHY-1419765), and by the JSPS KAKENHI (Grant Numbers 23740223 and 25287065). 
The numerical calculations were performed on SR16000 at the Yukawa Institute for Theoretical Physics, Kyoto University.

\bibliography{Garg_Mg24}

\end{document}